%% file: gallagher_ms.tex
\documentclass[usenatbib]{mn2e}
\usepackage{subfig}
\usepackage{graphicx}

\newcommand{\msun}{M$_{\odot}$}
\newcommand{\chandra}{{\emph {Chandra}}}
\newcommand{\hst}{{\emph {HST}}}
\newcommand{\cmsq}{{cm$^{-2}$}}
\newcommand{\aox}{$\alpha_{\rm ox}$}
\newcommand {\gtrsim} {\ {\raise-.5ex\hbox{$\buildrel>\over\sim$}}\ }
\newcommand {\lesssim} {\ {\raise-.5ex\hbox{$\buildrel<\over\sim$}}\ } 
\newcommand{\kms}{{km~s$^{-1}$}}

\title[The Windy Torus]{Investigating the Structure of the Windy Torus in Quasars}
\author[Gallagher et al.]{S. C. Gallagher$^{1,2}$\thanks{E-mail:
sgalla4@uwo.ca}, J. E. Everett$^{3}$, M.~M. Abado$^{1}$, and S.~K. Keating$^{4}$\\
$^{1}$University of Western Ontario, 1151 Richmond St, London, ON N6C 1T7, Canada\\
$^{2}$Visiting Fellow, Yale Center for Astronomy and Astrophysics, 
Department of Physics, Yale University\\
$^{3}$Center for Interdisciplinary Exploration and Research in Astrophysics (CIERA) and Department of Physics \& Astronomy, Northwestern University\\
$^{4}$Department of Astronomy, University of Toronto}

\begin{document}

\date{}

\pagerange{\pageref{firstpage}--\pageref{lastpage}} \pubyear{2014}

\maketitle

\label{firstpage}

\begin{abstract}

Thermal mid-infrared emission of quasars requires an obscuring
structure that can be modeled as a magneto-hydrodynamic wind in which
radiation pressure on dust shapes the outflow.  We have taken the
dusty wind models presented by Keating and collaborators { that
generated} quasar mid-infrared spectral energy distributions (SEDs),
and explored their properties (such as geometry, opening angle, and
ionic column densities) as a function of Eddington ratio and X-ray
weakness.  In addition, we present new models with a range of magnetic
field strengths and column densities of the dust-free shielding gas
interior to the dusty wind.  We find this family of models -- with
input parameters tuned to accurately match the observed mid-IR power
in quasar SEDs -- provides reasonable values of the Type~1 fraction of
quasars and the column densities of warm absorber gas, { though it
does not explain a purely luminosity-dependent covering fraction for either}.
Furthermore, we provide predictions of the cumulative distribution of
$E(B-V)$ values of quasars from extinction by the wind and the shape
of the wind as imaged in the mid-infrared.  Within the framework of
this model, we predict that the strength of the near-infrared bump
from hot dust emission will be correlated { primarily} with
$L/L_{\rm Edd}$ { rather than luminosity alone}, with scatter induced by
the distribution of magnetic field strengths.  The empirical successes
{ and shortcomings} of these models warrant further investigations
into the { composition and behaviour of dust} and the nature of
magnetic fields in the vicinity of actively accreting supermassive 
black holes.

\end{abstract}


\section{Introduction}

Researchers currently studying active galactic nuclei agree on a few
key premises.  The central engine is powered by a supermassive black
hole, and the optical through ultraviolet continuum emission is
generated by optically thick material in an accretion disk within the
central parsec down to a few gravitational radii.  Important specifics
such as the thickness of the disk as a function of radius are still
being resolved, but the overall picture is consistent; it is hard to
avoid disk formation given the persistence of angular momentum.
{ Within the central parsec}, ionized gas moving at high velocities covers
approximately $\sim10$\% of the sky and gives rise to broad
(1000s~km~s$^{-1}$) resonance and semi-forbidden emission lines that
characterize Type~1 quasars and Seyfert galaxies.  In many luminous
quasars, the broad UV emission lines show signatures of winds; the
most obvious of these are the P Cygni-type profiles seen in the high ionization
resonance lines of Broad Absorption Line (BAL) quasars
\citep[e.g.,][]{weymann+91}.  More subtle are the blue-shifts and
asymmetries seen in the same lines in other luminous quasars
\citep[e.g.,][]{richards+11}.  The wind manifested in these lines is
{ likely} driven by radiation pressure on ions at UV resonance
transitions.  For the wind to be launched in the vicinity of the
broad-line region, the quasar continuum must not only have significant
power in the UV -- the source of the line-driving photons -- but not
too much power in the X-ray.  X-rays will strip atoms such as C and O
of their electrons, thus dropping the absorption cross section
dramatically.  Overionized gas can therefore only be driven
radiatively by Thomson scattering, which is much less efficient than
line driving (e.g., \citealt{everett_ballantyne04}).  The elegant
picture of \citet{murray+95} that unifies the broad emission lines
seen in all Type~1 quasars with the BAL wind directly observed in only
$\sim20$\% of optically selected quasars (e.g.,
\citealt{hewett_foltz03}) by generating broad emission lines in an
equatorial disk wind is compelling, but has not been widely adopted in
other models. In such models, broad-line ``clouds'' are still invoked
as the source of the broad-line emission, though the smoothness of the
broad-line profiles and the requirement of some means of
pressure-confinement to prevent the putative clouds from shredding
remain persistently unsolved problems in the cloud picture (e.g.,
\citealt{murray_chiang97}; though see \citealt{bottorff_ferland00}).

An important part of active galactic nucleus (AGN) models that is even
more uncertain is the so-called torus, a dusty molecular region
presumably on the outskirts of the accretion disk.  The inner radius
of the torus is set by the dust sublimation radius, beyond which some
fraction of the accretion disk continuum is absorbed by grains and
reradiated.  Functionally, the torus must account for the $\sim30\%$
of the integrated radiant quasar power that comes out in the
near-to-mid-infrared.  Furthermore, it must obscure the central
continuum and broad-line region in a significant fraction of objects
called Type~2 AGN.  While a static toroidal structure \citep[whether
smooth or clumpy, e.g.,][]{kro_beg88,nenkova+02,nenkova+08} beyond the
dust sublimation radius will serve these purposes, such a cold
(100--1500~K) structure with the dust mass implied by the mid-infrared
luminosity would quickly become gravitationally unstable and collapse.
This problem was addressed by \citet{konigl_kartje94} with a
magneto-hydrodynamically launched wind, shaped also by radiation
pressure on dust.  { Proposing an alternate dynamical model}, \citet{elitzur_shlosman06} use
magnetic fields to launch and confine dusty clouds from the plane of
the accretion disk.  (\citealt{lawrence_elvis10} suggest an alternate
picture where warped disks, originating from gas in the plane of the
host galaxy, account for the large scale-height needed to obscure the
central engine; the gravitational instabilities that can give rise to
such asymmetries are further explored by \citealt{hopkins+12}.)  The
dusty wind picture was expanded by \citet[][hereafter K12]{keating+12}
by incorporating radiation driving on dust into the
magneto-hydrodynamic (MHD) wind models of \citet{everett05}, and then
illuminating the wind with the central continuum to generate a model
spectral energy distribution (SED).  While the details of the shape of
the output continuum still require some refining to match those
observed, the overall power and general shape generated by the K12
`fiducial' model is promising.  This model uses as inputs the
empirical SDSS composite quasar continuum \citep[with the
X-ray-to-optical flux ratio, $\alpha_{\rm ox}$, set to --1.6 as
appropriate for the input luminosity;][]{richards+06}, a black hole
mass of $M_{\rm BH}=10^{8}$~M$_\odot$, an Eddington ratio of $L/L_{\rm
Edd}=0.1$, and a column density at the base of the wind of $N_{\rm
H,0}=10^{25}$~cm$^{-2}$.  We further assume a Milky Way
interstellar-medium dust distribution \citep{draine_li07}.  In this
article, we expand on the results presented in K12 by exploring how
key input model properties affect the structure of the wind and
extracting other important observables that are generated by the wind
model.  Finally, we offer some predictions for observations that would
further test this picture.

\section{The Dusty Wind Model}

Following \citet{blandford_payne82} and \citet{konigl_kartje94}, we
model the torus as a self-similar dusty wind driven by MHD forces and
radiation pressure. The MHD radiative wind code has been described in
more detail previously \citep{everett05,keating+12}, and we give only a
brief outline here. First, a purely MHD-driven wind solution is
calculated given a set of initial conditions input by the user.  Next,
{\sc cloudy} (version 06.02.09b, as described by \citealt{Cloudy}) is called to
calculate the radiation field and dust opacity in the wind, allowing
for the model to determine the radiative acceleration of the wind due
to the central accretion disk continuum.  A new MHD solution is
calculated for the wind, now taking into account the radiative
acceleration, and the process is iterated until the wind converges to
a stable geometry.

Operationally, our assumption of self-similarity means that several
important quantities, including magnetic field strength and mass
density, scale with spherical radius; this simplifies the calculation
of the solution to the MHD equations considerably.  Furthermore, it
allows for the incorporation of radiation pressure from the disk by
simply reducing the effective gravitational potential. Importantly, it
allows us to use one streamline as an approximate template for the
entire wind, simply scaling appropriately inwards or outwards to find
properties of the specific region of the wind under consideration.  In
addition, these assumptions allow us to comparatively quickly estimate
the large-scale properties of these winds.  { The relevant input
parameters and physical properties of the specific models discussed
below are presented in Table~\ref{tab1}.}

\subsection{Density Distribution}

The final, converged MHD radiative wind model therefore provides the
mass density of the wind as a function of \emph{r}, the radial
distance from the axis of rotation of the disk, and \emph{z}, the
vertical distance from the plane of the disk. In more detail, the
model supplies the density along each streamline as a function of the
mass density at the base of the streamline and of polar angle. The
density of the wind at any arbitrary point is then relatively
straightforward to calculate as a consequence of the self-similarity
assumption. Because we know the shape of the wind streamlines, we can
use the shape of a single streamline to trace back from an arbitrary
point to the disk-launching radius for any wind streamline.  From
there, given a number density of particles on the disk at the inner
launch radius and the fact that density is assumed to be proportional
to $r^{-1.5}$ \citep{blandford_payne82}, we calculate the number
density at the launch radius of that particular streamline. To convert
from number density to mass density, we assign the proton mass to the
average particle.

\begin{figure}
\center
\includegraphics[width=\linewidth]{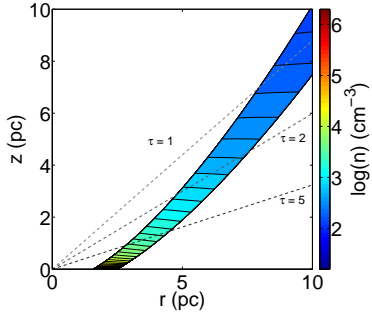} 
\caption{\label{fig:map} A 2D cross-section of the number density map of the fiducial dusty wind model  with $N_{\rm
H,0}=10^{25}$~cm$^{-2}$, $M_{\rm BH}=10^8$~M$_\odot$, and $L/L_{\rm
Edd}=0.1$ { (model 1 in Table \ref{tab1})}.  The color indicates number density according to the scale bar on the right. The number density drops sharply as a function of height,
$z$, as the wind accelerates.  Dashed straight lines indicate the values of the optical depth $\tau$ at H$\alpha\lambda6563$, as labeled.   
}
\end{figure}

In Figure~\ref{fig:map} we present a 2D color map of the number density
of one quadrant of the dusty wind as a function of \emph{r} and
\emph{z} for the fiducial model. The inner wind launch radius is set
by the dust sublimation radius. The notable thinness of the wind is a
consequence of the model parameters.  In general, the outer wind launch
radius is given by:
\begin{equation}
  r_{\rm outer} = r_{\rm in}\left(1-\frac{N_{\rm H, 0}}{2n_{\rm in}r_{\rm in}}\right)^{-2}
\label{eq: thickness}
\end{equation}
for user-specified values of the inner launch radius $r_{\rm in}$, the
number density at the inner launch radius $n_{\rm in}$, and the column
density along the base of the wind $N_{\rm H, 0}$.

%------------------------------------------------------------
%------------------------------------------------------------

\subsection{Defining the Covering Fraction}

Because the dust is entrained in the wind, $E(B-V)$ (calculated
explicitly by {\sc cloudy}) is a smooth function of inclination angle,
monotonically increasing toward the plane of the disk where the polar
angle $\theta=90^\circ$. We see this continuous distribution of
density as a strength of the dynamical model presented, however, these
values offer no obvious inclination angle to choose as the boundary
for the wind, and the number density map does not show any obvious
feature that would define a `height' for the torus.
%Figure~\ref{fig:tau} shows the optical depth as a function of polar
%angle $\theta$ for a few selected wavelengths. 
{ The optical depth through the wind 
is also a smooth function of $\theta$ for a wide range of
wavelengths, and similarly provides no obvious angle to mark the
boundary of the wind.}  We have therefore chosen the angle at which the
dust reaches an optical depth of 5 for light at a wavelength of
6563~\AA\ (the wavelength of H$\alpha$ emission) to be a { conservative}
definition of the opening angle (i.e., the inclination angle up to
which the central engine may be detectable { with high signal-to-noise ratio spectra} to an observer) of the
dusty wind. At this value of $\tau$, only 0.7\% of the broad H$\alpha$
emission is transmitted through the wind, and so a quasar viewed from
this inclination angle (or closer to the disk) would be classified as
an optical narrow-line object.  For reference, this value of
$\tau_{\rm H\alpha}$ corresponds to $E(B-V)\approx1.6$ and $A_{\rm V}\approx5.5$ ({\sc cloudy} assumes $R_{\rm V}=3.4$).  For our
  fiducial model, $\theta_{\rm open}=69^{\circ}$; { the opening
  angles for all of the models are listed in Table~\ref{tab1}. 
A value of $\theta_{\rm open}=69^{\circ}$ gives a dusty wind covering fraction of $\cos(\theta_{\rm open})=$35\%.  As expected, this is approximately the fraction of the quasar bolometric luminosity that is emitted from 2--100\micron\ in the predicted SEDs from K12.  For completeness, we also include the less restrictive opening angle
  values for $\tau_{\rm H\alpha}=3$ (corresponding to 5\% transmission
  of the H$\alpha$ line flux and $A_{\rm V}\approx 3.3$) which is
  more relevant for low signal-to-noise ratio optical spectra typical
  of extragalactic survey data.}

\section{Dependence of the Wind on Input Parameters}

\subsection{The Effect of \boldmath$L/L_{\rm Edd}$}

\begin{figure}
\center
\includegraphics[width=\linewidth]{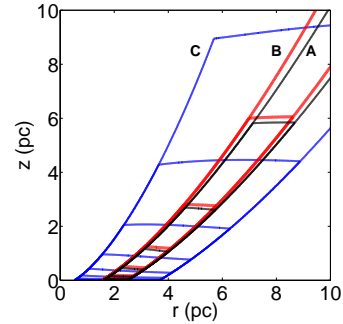} ~
\caption{\label{fig:streamlines} A cross section of the wind, as in Figure~\ref{fig:map},
  comparing the profiles of several wind models.  Shown here is the
  fiducial model (black, labelled ``A''{ ; model 1 in Table~\ref{tab1}}), with $N_{\rm
    H,0}=10^{25}~\rm{cm}^{-2}$, $L/L_{\rm Edd}=0.1$, and $M_{\rm
    BH}=10^8~\rm{M}_{\odot}$; a similar
  model with $M_{\rm BH}=5\times10^8~\rm{M}_{\odot}$ (red, ``B''; { model 2 in Table~\ref{tab1}}); and a third model
  with the same parameters as the fiducial model except that $L/L_{\rm Edd}=0.01$ (blue, ``C''; { model 3 in Table~\ref{tab1}}). { The dimensions of the model B streamlines have been scaled by $1/\sqrt{5}$ to directly compare the shape of the outflow to the other models.} 
Increasing the black hole mass by a
  factor of 5 (keeping $L/L_{\rm Edd}$ constant) does not significantly change the shape of the wind,
  but reducing $L/L_{\rm Edd}$ by a factor of 10 moves the inner launch radius in by a factor of $\sqrt{10}$ (as expected), and causes the inner wind streamlines to be more vertical.  This would have the effect of reducing the projected cross-section of the base of the inner wind (which is optically thick to near-IR emission; { see Fig. 7 in} K12) compared to what is seen in the fiducial model.  We propose this geometric explanation for the increased prominence of the 3--5~\micron\ bump in luminous quasars. 
}
\end{figure}

Within the context of our model, we can explore variations in windy
torus geometries across the AGN population by investigating the
effects of changing luminosity, $L/L_{\rm Edd}$, and $M_{\rm BH}$ on
the structure of the wind.  Figure~\ref{fig:streamlines} illustrates
the shape of the fiducial wind model (labeled as A{; model 1 in Table~\ref{tab1}}) compared to two
other models. Model B (model 2 in Table~\ref{tab1}) has a black hole mass of
$5\times10^8~M_{\odot}$, five times that of the fiducial model.
Because the value of $L/L_{\rm Edd}$ is fixed at 0.1, this model is 5
times more luminous than the fiducial model.  The resulting shape of
the wind is nearly identical to that of the fiducial model; the
streamlines are slightly more vertical.  Model C { (model 3 in Table~\ref{tab1})} has a black hole mass
of $10^8$~\msun -- the same as the fiducial model -- but with
$L/L_{\rm Edd}$ dialed down to 0.01.  The inner launch radius moves in
by a factor of $\sqrt{10}$, as expected with an $L^{0.5}$ scaling of
$r_{\rm in}$, and the inner streamlines are more vertical as radiation
pressure is relatively less important than at higher Eddington ratios.
The slight change in wind structure from altering $M_{\rm BH}$ but
keeping $L/L_{\rm Edd}$ constant compared with the larger change from
decreasing $L/L_{\rm Edd}$ by a factor of 10 shows that the Eddington
ratio is a primary driver given our modelled wind properties.

%------------------------------------------------------------
%------------------------------------------------------------

\subsection{The Effect of \boldmath{$B-$}field Strength}

The organized, poloidal magnetic fields in our fiducial MHD-wind model
are reasonably strong, with mass-to-magnetic flux ratios of
$\kappa=0.03$, implying considerable magnetic fields strengths of
$\sim0.235$~G at the innermost streamline.
While an ordered magnetic field is required at the base of the wind to
lift material into the line of sight of the accretion disk continuum,
radiation pressure on its own may effectively accelerate material
after that point.  We explored the effect of dialing down the strength
of the magnetic field.  Because $\kappa\propto\frac{1}{B}$, 
this means
increasing $\kappa$, and so we ran models with $\kappa$=0.04,
0.06, and 0.12 { for values of $L/L_{\rm Edd}=0.10$ (models 4--6 in Table~\ref{tab1})}.  Changing $B$ to be up to 4$\times$ weaker had a
significant effect on the distribution of $E(B-V)$ values over the
sky.  We found that stronger fields (smaller values of $\kappa$) result in higher
$E(B-V)$ values over a larger fraction of the sky; the Type~2 boundary
for relatively stronger fields occurs at lower inclination angles
({ $69^\circ$ vs. $74^\circ$} for $\kappa=0.03$ and 0.12, respectively).
These results are a direct consequence of radiation driving dominating
when the fields are weaker, thus leading to more radial outflows.

\input{tab1}

We also ran the different field-strength models for lower $L/L_{\rm Edd}$
values of 0.05 (vs. 0.10 for the fiducial model), and found little
effect on the cumulative $E(B-V)$ distribution (see
Fig.~{\ref{fig:cum_dist}; { see models 7--9 in Table~\ref{tab1})}.  The lower $L/L_{\rm Edd}$ values also lead
  to smaller $r_{\rm sub}$ values (and
  thus $r_{\rm in}$ values for the wind), but the $E(B-V)$
  distribution appears to respond most sensitively to increasing
  $\kappa$ { because} radiation pressure { consequently} becomes more dominant.  At 
  mass-to-magnetic flux ratios of $\kappa\sim1$ the MHD wind will fail,
  and without an alterate source
  of vertical lift (e.g., infrared radiation pressure;
  \citealt{dorod+12}) radiative driving will also not be effective 
because material from the outer disk will not be exposed to the light from the inner disk.

%------------------------------------------------------------
\begin{figure}
\centering
\includegraphics[width=\linewidth]{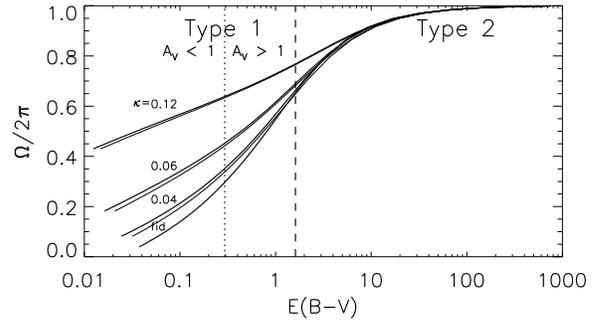} ~
\caption{\label{fig:cum_dist} 
The cumulative covering fraction of the wind ($\Omega/2\pi$)
vs. extinction, $E(B-V)$; larger $\Omega/2\pi$ values at low extinction show winds that cover less of the sky with significant extinction.  The dotted vertical line marks $A_V
= 1$, and the dashed vertical line marks the Type~1/Type~2 boundary as
defined by $\tau_{H\alpha}=5$.  The mass-to-magnetic flux ratios,
$\kappa$, for each model are as labelled; with thick (thin) curves
representing $L/L_{\rm Edd}=0.1$ (0.05).  The fiducial model curve
(marked `fid') has $\kappa=0.03$.  { The curves plotted are for models 1 and 4--9 in Table~\ref{tab1}.} Decreasing the strength of the
magnetic field { (increasing $\kappa$)} 
increases the importance of radiative driving and makes
the wind more radial; consequently, the opening angle of the wind
increases and the extinction to Type~1 objects decreases overall.}
\end{figure}

%------------------------------------------------------------

\subsection{The Effect of X-ray Weakness}

Within the context of the dusty wind model, we are also interested in
exploring the effect of the well-known anti-correlation between
$\alpha_{\rm ox}$ (the 2~keV to 2500~\AA\ spectral index) and $L_{\rm
UV}$ whereby more UV-luminous quasars emit a smaller fraction of their
power in the X-rays.  For our fiducial model, the value of
$\alpha_{\rm ox}$ is $-1.6$; we have also run models assuming
X-ray-loud ($\alpha_{\rm ox}=-1.1$) and X-ray weak ($\alpha_{\rm
ox}=-2.1$) versions of the illuminating SED.  Figure~\ref{fig:open}
shows the opening angle, $\theta_{\rm open}$, as a function of inner
radius, $r_{\rm in}$. For this large range of $\alpha_{\rm ox}$ values
(--2.1 to --1.1 encompasses the extremes of the observed
distribution), the effect of X-ray loudness on the opening angle is
not large, although two of the X-ray louder models do have slightly
larger opening angles ($\theta\sim75^{\circ}$) compared to the X-ray
quiet ones ($\theta\lesssim73$).  This result is reasonable, because
unlike the ionization parameter of the associated gas, dust driving
depends primarily on the integrated {\em power} in the optical through
X-ray, and less on the spectral {\em shape}.

The models with altered \aox\ presented in K12 were not particularly
realistic, as the X-ray continuum was modeled simply as a flat (in
units of power vs. frequency) continuum above 0.29~keV whose
normalization was set by \aox.  In an effort to create a more
realistic continuum, we instead include a dust-free layer of shielding
gas interior to the dusty wind.  The shape of the ionizing continuum
impinging on the dusty wind is therefore altered by gas absorption
more realistically.  To explore the effect of { increasing} the shielding gas { thickness}, we
ran four additional versions of the fiducial model where only the
column density of shielding gas has been altered with values of
$\log(N_{\rm H, shield})=19, 23, 24,$ and 25~\cmsq\ { (models 10--13 in Table~\ref{tab1}).  The fiducial
model has $\log(N_{\rm H, shield})$=18~\cmsq.}  When we calculate the
cumulative distribution of $E(B-V)$ as in Figure~{\ref{fig:cum_dist}},
we find that increasing the dust-free shield makes no difference in
the shape of that curve, and the opening angles for the dusty wind
remain the same as for the fiducial model.  Therefore, the shielding
gas does not strongly impact the covering fraction of the dusty-wind
torus, consistent with the results from our X-ray weak continuum
modeling.

\section{Observational Consequences of the Dusty Wind Model}

The geometric structure, dust distribution, column density, and
ionization state of material in the dusty wind model have direct
observational consequences.  In K12, we predicted
mid-IR SEDs of the wind.  Now, we are specifically interested in
exploring if the dusty MHD wind can account for the ratio of Type~1 to
Type~2 AGNs, the gross properties and prevalence of warm absorbers,
and the near-IR bump from hot dust at $\sim1200$~K in luminous AGNs.

\subsection{Broad vs. Narrow Line AGN in the Dusty Wind Picture}

Given the requirement that a successful torus model obscure the
accretion-disk continuum and broad-line region emission, comparing our
predicted fraction of Type~1 (optical broad-line) and Type~2 (optical
narrow-line) quasars to observations is an important test of the
models.  { In our models, the range of opening angles is relatively
tight, and shows little dependence on model input parameters with the
exception of the column density at the base of the wind (only models
with $N_{\rm H,0}=10^{25}$~\cmsq\ are shown in Fig.~\ref{fig:open}),
and the strength of the magnetic field (see Figs.~\ref{fig:cum_dist}
and \ref{fig:open}).}  Specifically for the suite of models run by
K12, typical opening angles using the $\tau_{\rm H\alpha}=5$ criterion
are $\theta_{\rm open}=69$--75$^\circ$ (see~Fig.~\ref{fig:open}); {
this corresponds to dusty-wind covering fractions (Type~2 fractions)
of $\frac{\Omega}{2\pi}=\cos(\theta_{\rm open})$=26--36\% or Type~1 to
Type~2 ratios of 1.8 to 2.8.  In their survey of low-redshift,
emission-line galaxies from the SDSS, \citet{hao+05} found that Type~1
objects outnumber Type~2 objects by a factor of 2--4 at high
luminosities.  A similar study by \citet{simpson05} found Type~1
fractions of $\sim60$\% (Type~1/Type~2 = 1.5) at the luminous end (in
terms of [O~{\sc iii}] luminosity).  Our numbers are consistent with
(though narrower than) the range of 1.5 to 4 in these population studies.}

{ As demographic studies can be plagued by selection effects, an
alternate means of constraining the covering fraction of the dusty
``torus'' is to systematically measure the power in the IR-region of
Type~1 quasar SEDs.  This assumes that the dusty medium acts
effectively as an efficient bolometer whose emitted power is simply
the fraction of the sky covered by the dusty medium from the point of
view of the accretion disk.  While radiative transfer modeling shows
this is not strictly true (cf. K12), it is a reasonable approximation.
Along these lines, the recent analysis by \citet{roseboom+13} using
{\em WISE} observations of luminous ($\sim10^{46}$~erg~s$^{-1}$)
Type~1 SDSS quasar determined that the covering fractions could be
described by a log normal distribution with mean of
$39^{+23}_{-14}\%$.  This correponds to an intrinsic Type~1/Type~2
ratio ranging from 0.61 to 3.0; the range from our suite of models of
1.8 to 2.8 is again within (though narrower than) the span of
empirical values.}

{ A Type~2 fraction that decreases with increasing luminosity does
not follow simply from our model in the parameter range we have explored, 
as the wind covering fraction is not
sensitive to only luminosity, but a combination of $L/L_{\rm Edd}$ and
the mass-to-flux ratio of the magnetic field.}  { However,} it
should be kept in mind that there can be other explanations for a
luminosity dependence for obscuration.  For example, the host galaxy
can also serve as a significant source of obscuration of the
broad-line region \citep[e.g.,][]{rigby+06,juneau+11}.  If accretion
disks are randomly oriented with respect to their host galaxies as
observed, the observed Type~2 fractions are inflated with respect to
the torus-covering fraction because of line-of-sight obscuration
occurring from optically thick dust within the host galaxy {
\citep{kinney+00}.}  Because the size of the BLR scales with
luminosity \citep[$\propto L^{0.5}$;][]{kaspi_rm,bentz+09}, more
luminous broad line regions are larger and therefore harder to hide,
even before invoking any sort of feedback mechanism that might clear
the host galaxy ISM from the vicinity of the quasar.  Furthermore,
weak broad lines from low-luminosity objects can be ``hidden'' by a
strong stellar continuum \citep{moran+02,martini+06}; detecting these
requires high S/N spectra and careful modeling and subtracting of the
stellar galaxy emission \citep[e.g.,][]{ho+97,hao+05}.

%------------------------------------------------------------
\begin{figure}
\center
\includegraphics[width=\linewidth]{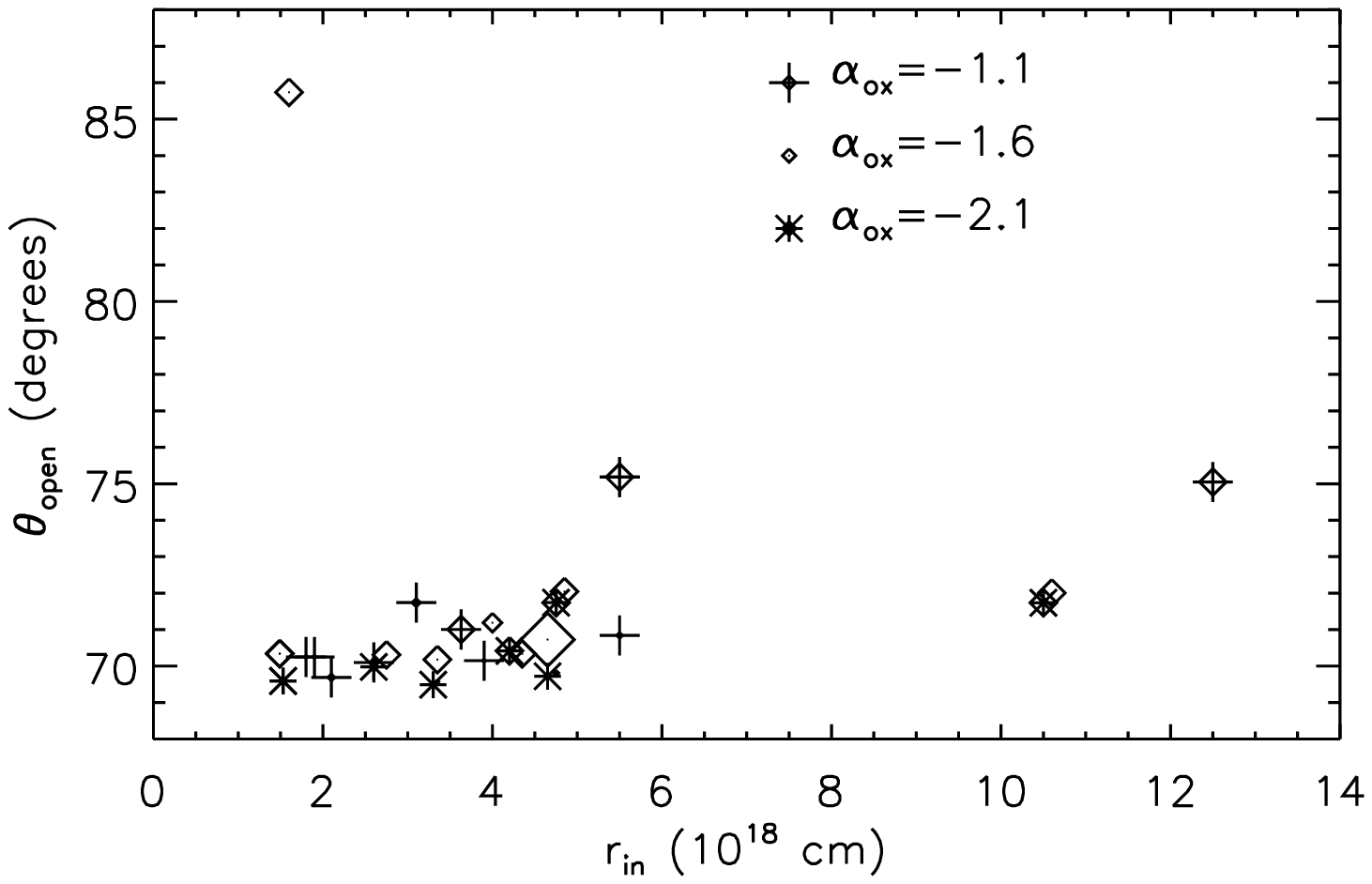}\\
\includegraphics[width=\linewidth]{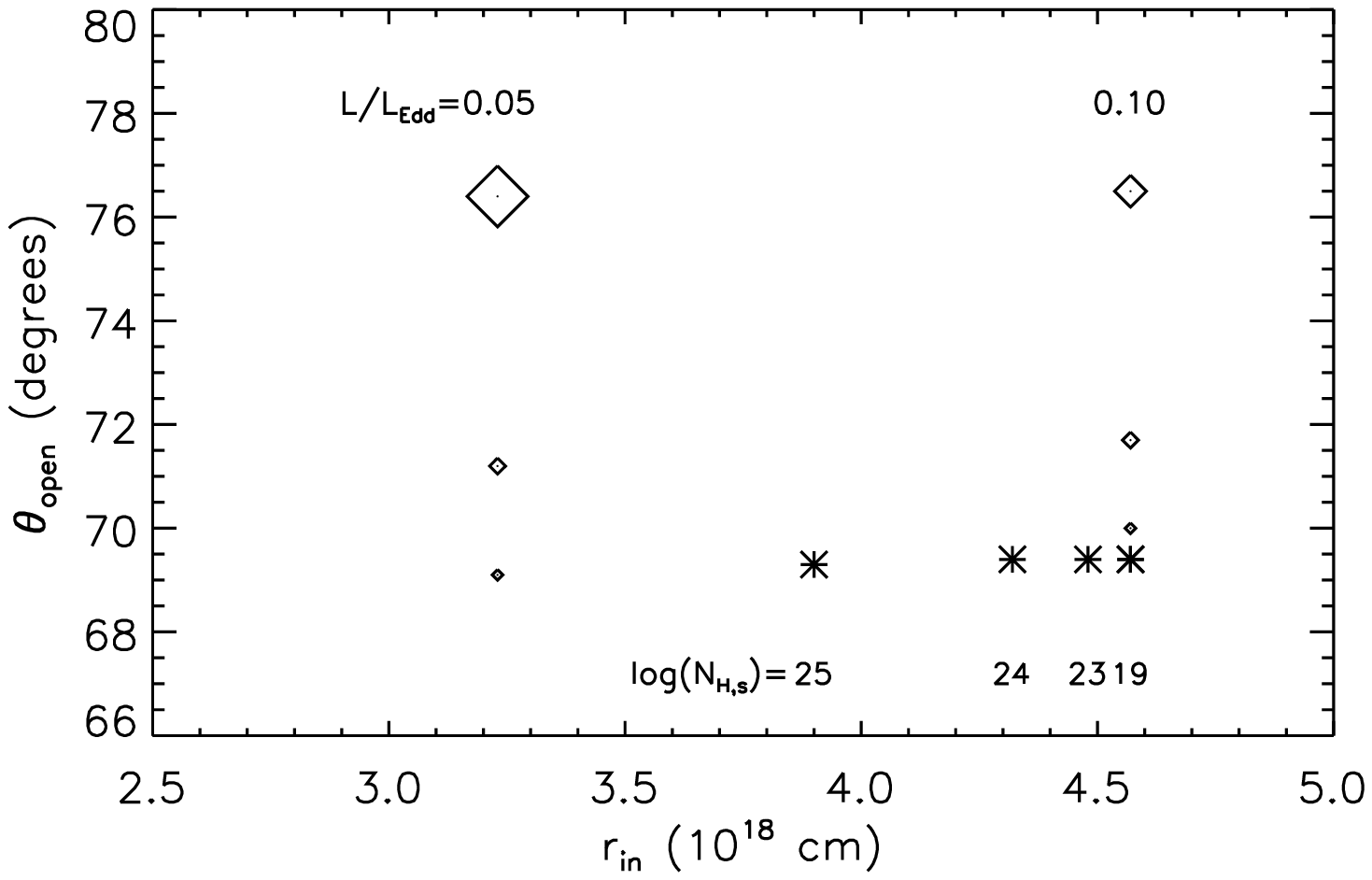} ~
\caption{\label{fig:open} 
{\em Top}: The opening angle, $\theta_{\rm open}$ (degrees), as a 
function of the
wind launch radius, $r_{\rm in}$ ($10^{18}$~cm).  All models have base
column densities of $N_{\rm H,0}=10^{25}$~cm$^{-2}$, the value of
$N_{\rm H,0}$ that generates appropriate near-to-mid-infrared power in
the SED \citep{keating+12}.  The crosses, diamonds and asterisks
indicate values of $\alpha_{\rm ox}$ as labeled in the legend.  The
diamonds are scaled to represent their $L/L_{\rm Edd}$ values; the
size shown in the legend is $L/L_{\rm Edd}=0.05$.  The models are
those presented in K12; the bulk of them give opening
angles of 69$^\circ$--75$^\circ$. 
 The outlier with $\alpha_{\rm ox}=-1.6$ at
$\theta_{\rm open}\approx86^{\circ}$ is the model with $r_{\rm in}$
set artificially low to mimic the effect of increasing the dust
sublimation temperature.  {\em Bottom}: Same as panel above for the new 
models with variable mass-to-flux ratios and variable shielding gas column densities.  Diamonds are scaled to represent $\kappa$ values from 0.03 
(smallest, i.e., strongest magnetic field strength) 
to 0.12 (largest; weakest magnetic field strength).  
Asterisks show $\kappa=0.03$ models 
with a range of shield values ($\log(N_{\rm H,shield})$ [\cmsq]) as labeled; all other models have $\log(N_{\rm H,shield})=18$ \cmsq.  
While decreasing the strength of the magnetic field increases 
$\theta_{\rm open}$, increasing the column density of the shielding 
gas has little effect on $\theta_{\rm open}$, although the launching radius decreases  with increasing $N_{\rm H,shield}$.}
\end{figure}
%------------------------------------------------------------

Here we are defining Type~1/Type~2 as broad-line/narrow-line with
respect to H$\alpha$; however, the X-ray community often uses the
detection of intrinsic X-ray absorption with $N_{\rm H} \ge
10^{22}$~\cmsq\ to distinguish obscured vs. unobscured quasars.  The
two criteria overlap, but do not define identical populations, as dust
extinction and gas absorption do not always go hand-in-hand.  The
$N_{\rm H}$ value at what we have defined as our Type~1/2 boundary is
$\sim10^{22}$~\cmsq, and therefore consistent with both the X-ray and
optical definitions.  In the wind model, this also sets a floor for
the amount of intrinsic absorption that would be found in optical
Type~2 objects { where the dusty wind is obscuring the broad line
region}, consistent with the Seyfert~2 X-ray absorption study of
\citet{risaliti+99} and observations of X-ray occultation events by
\citet{mark+13}.

\subsection{Ionic Absorption Through the Wind}

\subsubsection{Warm Absorbers}

As seen in Figure~\ref{fig:map}, wind density drops quickly as a
function of height above the plane of the accretion disk.  Because the
wind is relatively collimated and narrow radially, this means that the
line-of-sight column density also drops rapidly with decreasing
inclination angle.  The wind at moderate
inclination angles is therefore exposed to the high-energy continuum
from the accretion disk and becomes highly ionized.  Such highly
ionized gas along the line-of-sight to the X-ray continuum is observed
in many AGN ($\sim 50\%$ of Seyfert galaxies;
e.g., \citealt{reynolds97}); the strongest atomic features of these
so-called ``warm absorbers'' are the bound-free edges for O~{\sc vii}
and O~{\sc viii} and the helium-like and hydrogen-like inner shell
bound-bound transitions of these ions.  

Studies of the variability of X-ray absorption features have led to
loose constraints on the location of warm absorbers that are largely
consistent with the torus location \citep[e.g., within
$\sim6$~pc;][]{krongold+05}.  The specific identification of the warm
absorber with a hot phase within the vicinity of a molecular torus has
been explored previously.  Indeed, this was suggested by
\citet{reynolds97} when {\em ASCA} observations revealed the
prevalence of warm absorbers in Seyfert galaxies 
{ with some preference for those with heavier optical extinction.}  
Subsequently, \citet{krolik_kriss01} explored a multi-temperature thermal-wind
origin for the warm absorber gas: they postulated that warm absorbers
were evaporated off of the inner surface of the dusty torus.

More recently, \citet{dorod+12} presented a radiation-hydrodynamic
model that incorporated infrared radiation pressure on dust to drive a
slow wind (10--200~\kms) off of the accretion disk beyond the dust
sublimation radius.  The inner (from the point of view of the
accretion disk) layer of this slow, cold wind is illuminated by a
bright X-ray continuum to form a hot, fast (400--800~\kms) wind.  In
the parameter space they considered, UV radiation pressure on dust
grains was not included, and the X-ray luminosity needed to power the
wind was extremely high, up to $L_{\rm X}=0.3L_{\rm Edd}$, much
larger than seen in luminous quasars (where $L_{\rm X}\sim0.01L_{\rm
Edd}$ is more typical).  The opening angle for the Thomson-thick
part of the torus was $72^\circ-75^\circ$, which is considerably
smaller than our values (close to $90^{\circ}$).

With high S/N X-ray gratings spectra, O~{\sc vii} and O~{\sc viii}
lines and edges in absorption can be used to constrain the column
densities of the X-ray absorbing oxygen ions.  For reference, two
high-quality examples of this type of analysis are based on the 900~ks
\chandra\ observation of NGC~3783 \citep{kaspi+00} and the
simultaneous \hst/\chandra\ study of Mrk~279
\citep{costantini+07,arav+07}; both Seyferts are X-ray bright with
particularly strong warm absorber features.  The best-fitting column
densities from these studies range from 1--$5\times10^{17}$~\cmsq\
(O~{\sc vii}) and 0.3--$39\times10^{17}$ (O~{\sc viii}).  The more
recent X-ray study of the reddened quasar IRAS~13349+2438 by
\citet{lee+13} found similarly high column densities of
$10^{17-18}$~\cmsq\ for both ions.  From the wealth of features in the
HETGS spectrum, the authors favoured a model with an absorber with a
continuous distribution of ionization parameters, and suggested that
it could be an ionized atmosphere above the putative torus as in the
\citet{krolik_kriss01} picture.

The column densities from these detailed case studies of the strongest
{ (and therefore -- by selection -- not typical)} warm absorbers
are about an order of magnitude higher than we find in our fiducial
model for O~{\sc viii}, and about a factor of 100 larger that our
values for O~{\sc vii} (see Fig.~{\ref{fig:columns}).  Our X-ray weak
modification to the fiducial model (with $\alpha_{\rm ox}=-2.1$)
however, gives values of O~{\sc viii} and ~{\sc vii} that are both
$\sim10^{17}$~\cmsq\ over a large range of inclination angles.  The
X-ray loud ($\alpha_{\rm ox}=-1.1$) fiducial model predicts much lower
values for the O~{\sc viii} and O~{\sc vii} column densities because
most of the oxygen nuclei are completely stripped of electrons.  While
the distributions for the fiducial models with the X-ray modifications
are all quite flat as a function of inclination angle for
$\theta<\theta_{\rm open}$, the $\kappa=0.12$ model (with the weakest
$B$-field), has a much smaller range of angles over which the ``warm
absorber'' gas would be detected (from
{ $\theta=55^{\circ}-76^{\circ}$}), but the hydrogen-like and helium-like
oxygen column densities rise steeply as the inclination angle
approaches $\theta_{\rm open}$, and near that boundary have values
consistent with the empirical values measured from { the extreme}
X-ray spectra cited above (see~Fig.~\ref{fig:columns}).  Adding a
layer of { dust-free} shielding gas up to $N_{\rm
H,shield}=10^{25}$~\cmsq\ to the fiducial model at most approximately
doubles the O~{\sc vii} and O~{\sc viii} column densities.  Given the
range of values obtained from the models, we consider the high
ionization gas within the MHD+radiation-driven wind a plausible
explanation for { more typical (lower column density)} 
X-ray warm absorbers.  { All of our models by
design have ``dusty warm absorbers''; the dust
 is collisionally coupled to the gas which is launched and collimated 
by magnetic fields to large scale heights where it can be exposed to the
 X-ray continuum and cover a large fraction of the sky.  The model in
 its present form checks (via {\sc cloudy}) that the gas temperature
 is consistent with dust survival along the entirety of a streamline
 to determine the innermost boundary of the wind.  
Nonetheless, incorporating the data from Figs. {\ref{fig:cum_dist}} and {\ref{fig:columns}}a, there is a large range of inclination angles ($\theta=15$--$45^\circ$) in the fiducial model 
with a substantial warm absorber column density of $N_{\rm OVIII}\gtrsim10^{16}$~\cmsq\ without severe reddening ($A_{\rm V}<1$).  The models with weaker $B$-fields (see Fig.~\ref{fig:columns}b) generate dusty warm absorbers exclusively.}

It is not obvious how one could distinguish observationally between
the different proposed scenarios for the warm absorber origin with
present data, but future missions such as {\em Athena+} that will
better detect and resolve the velocity structure of X-ray absorption
features { (including dust signatures; e.g., \citealt{lee+05})}
may shed some light on the relationship of the warm absorbers to
a dusty wind.  Our models currently provide only integrated column
densities of lines-of-sight through the dusty wind, and so we cannot
directly compare absorption profiles as a function of outflow velocity
with observed spectra in the UV and X-ray.  To provide some reference,
the maximum vertical outflow velocities are 
a few $10^{3}$~\kms\ { (see Table~\ref{tab1})}.  
Warm absorbers typically have multiple velocity
components present that can vary independently, and the radiatively
driven dusty wind could plausibly account for some of them.  However,
explaining the details of individual X-ray spectra frequently requires
some sort of inhomogeneous structure, and we do not claim that the
MHD-wind in this iteration of our model is sufficient to account for
all of the spectral features.  Nonetheless, the presence of highly 
ionized gas
in the dusty outflow occurs naturally within the MHD plus UV/optical
and X-ray radiation-pressure driven model.

\subsubsection{Low Ionization Gas}

If we use Mg~{\sc ii} as a tracer of low ionization gas, it is
apparent from Figure~{\ref{fig:columns}} that significant column
densities of this ion are not visible in Type~1 objects; the column
density of Mg~{\sc ii} only rises steeply at inclination angles larger
than the opening angle.  In this case, the dusty wind alone cannot
give rise to the low ionization gas responsible for broad Mg~{\sc ii}
absorption lines seen in LoBAL quasars.  However, in the fiducial
model, we have not included a layer of dust-free shielding gas or the
UV BAL outflow, as would of course be required for a robust model of
broad absorption line quasars.  Along the line-of-sight from the X-ray
continuum from the inner accretion disk, BAL quasars have column
densities of at least $10^{22}$~\cmsq\ and low-ionization BAL quasars
may be either Compton-thick \citep[e.g.,][]{gall+06} or intrinsically
X-ray weak \citep[e.g.,][]{luo+13} with observed \aox\ values of
$\lesssim-2.0$.  (Our addition of MHD-driven shielding gas for some
model runs does not include UV radiation-driving, and so it is not a
realistic model for a BAL quasar.)

Relatively rare among optically selected quasars, LoBAL quasars have
dust-reddened continua that are intrinsically quite blue
\citep{sprayberry_foltz,reichard+03}, and may be evolutionarily young
systems \citep[e.g.,][]{farrah+05,urrutia+09}.  Alternatively, all BAL
quasars may have a LoBAL region that is only probed through some
fraction of the lines of sight through the wind.  This is expected in
a radiatively driven wind launched from an accretion disk given the
stratified structure of the broad line region where higher ionization
gas (probed by C~{\sc iv}) originates at smaller radii than lower
ionization gas (probed by Mg~{\sc ii}).  The similarity of the high
ionization and low-ionization BAL velocity structure (at least at low
velocities; \citealt{voit+93}) is consistent with both features being
found within the same structure launched from a range of radii close
to the source of the UV-optical continuum emission and well within the
dust sublimation radius.  The dust extinction which is common in LoBAL
quasars may however arise from a dust-driven wind as presented in this
paper.  The concurrence of extinction with LoBALs in a single object
therefore would not indicate that there is necessarily dust entrained
in the LoBAL outflow; the large inclination angles required to pass
through the LoBAL gas are also consistent with larger values of
$E(B-V)$ from the dusty wind.

%-----------------------------------------------------------
\begin{figure}
\centering
\includegraphics[width=\linewidth]{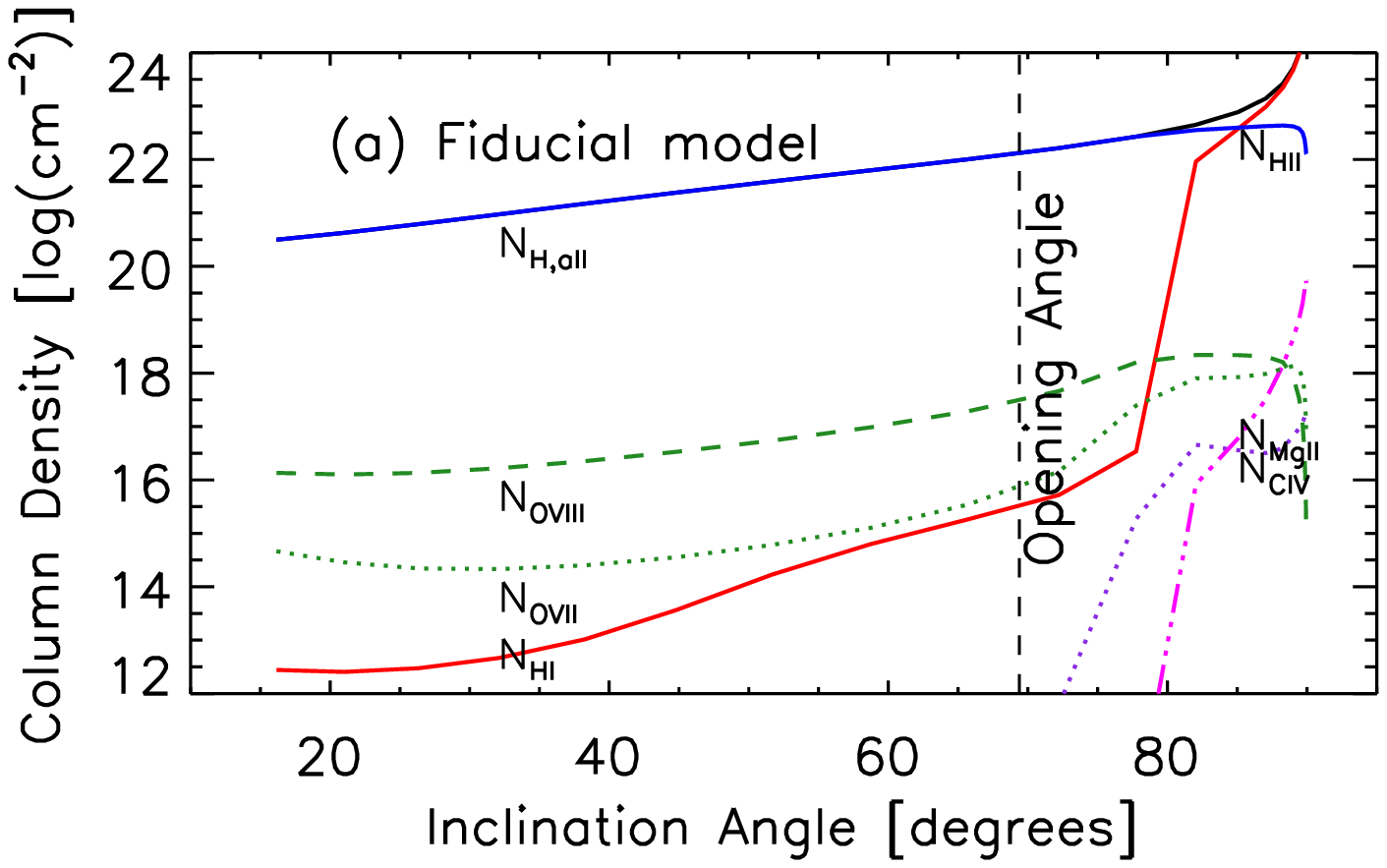}\\
\includegraphics[width=\linewidth]{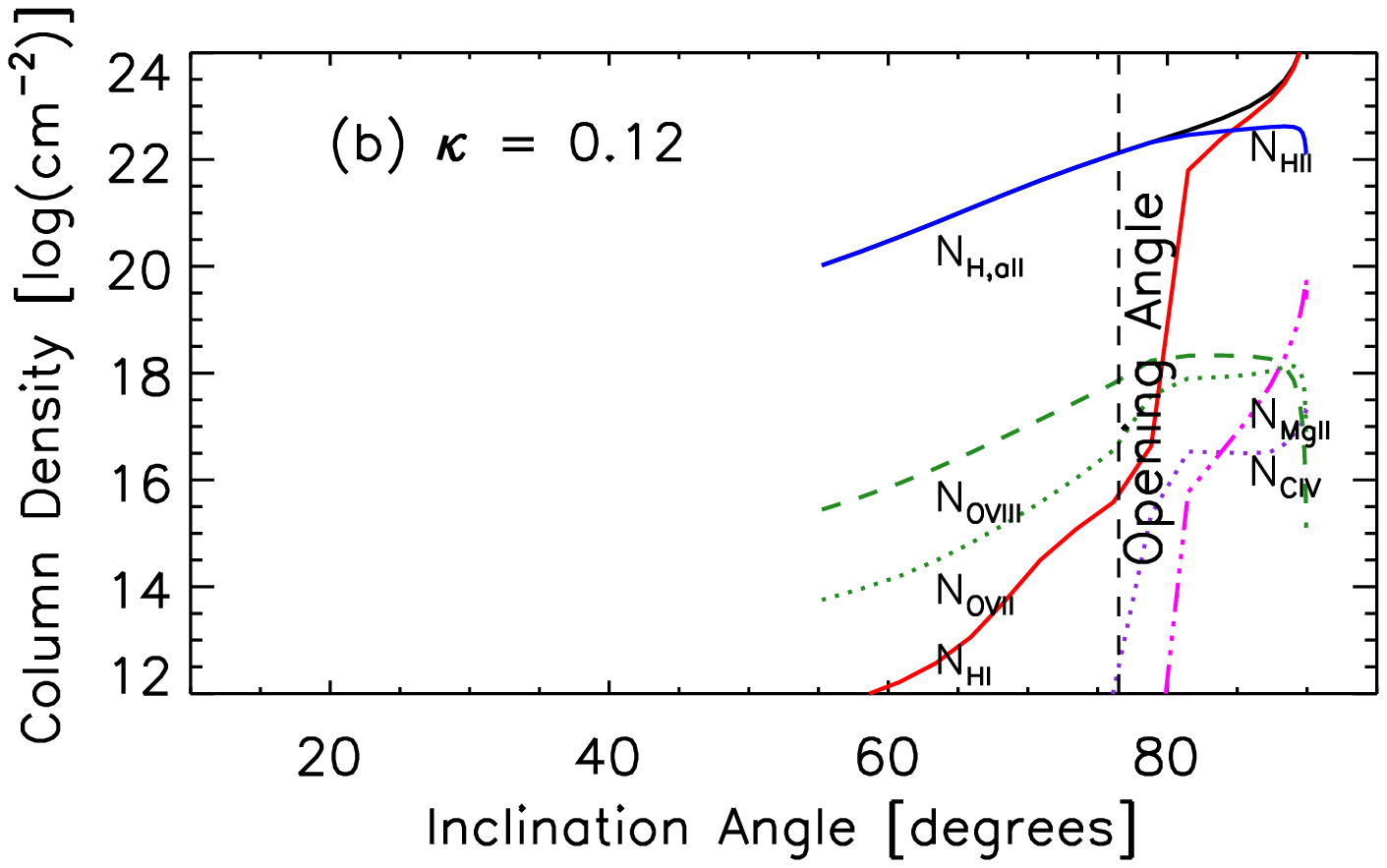}~
\caption{\label{fig:columns} {\em Top} (a):
  Column density vs. inclination angle of abundant ions
  characteristic of both high ionization (e.g., O~{\sc vii} and O~{\sc
    viii}) and low-ionization gas (e.g., H~{\sc i} and Mg~{\sc ii})
    for the fiducial K12 model.  {\em Bottom} (b): The same plot as
    above for the lowest $B$-field model with $\kappa=0.12$.  Because
    of the increased importance of radiative driving, the ionized gas
    covers a significantly smaller fraction of the sky in this model,
    and the O~{\sc vii} and O~{\sc viii} column densities increase
    more steeply and to larger values than in the fiducial model with
    $\kappa=0.03$.  In both panels, the column densities of
    high-ionization gas are largely consistent with constraints from
    warm absorber studies.
    The low-ionization species
  have negligible column densities at inclination angles smaller than
  the opening angle (marked with a vertical dashed line).
}
\end{figure}

%------------------------------------------------------------

\subsection{Hot Grains in Dusty Winds}

Within the context of the wind model, we would also like to investigate
the empirical observation that the strength of the 3--5~$\mu$m bump --
the manifestation of a larger contribution to the near-to-mid-infrared
SED from hot dust -- is seen to increase with increasing quasar
luminosity { (e.g., \citealt{edelson_malkan, richards+06,gall+07,
  coleman+13,roseboom+13})}.  In our picture, the inner streamline carries the
hottest dust, and it becomes more visible as $L/L_{\rm Edd}$ increases
because the relative importance of radiative driving goes up and the
wind is pushed closer to the equatorial plane { \citep[][K12]{gall+07}}.  We
predict that this would have the effect of increasing the prominence
of the 3--5~\micron\ emission, as the projected cross section of the
surface of the inner streamline from a typical Type~1 orientation
angle will increase.  The dusty wind becomes optically thick to
near-infrared emission at large inclination angles { (see Fig. 7 in K12)}.
Consequently, within the dusty wind paradigm, we expect that the
claimed correlation of the strength of the 3--5~\micron\ emission with
luminosity is in fact a correlation with Eddington ratio.  (For
luminous quasars, luminosity and $L/L_{\rm Edd}$ are of course
correlated given the restricted range of observed black hole masses in
flux-limited quasar surveys; e.g., \citealt{vestergaard_osmer09}.)

%graphite 

A second means of boosting the 3--5~\micron\ emission is to decrease
the inner launch radius in the fiducial model by about one-third by
increasing the dust sublimation temperature to approximate the effect
of a graphite-driven wind (K12).  This decrease in the inner launch
radius causes the wind to be notably broader in our model similar to
what we saw with the decrease in $L/L_{\rm Edd}$ and as expected from
Equation~\ref{eq: thickness}.  However, the {\em shape} of the
streamlines is very close to that of Models A and B.  Because the bulk
of the near-infrared emission is coming from the innermost
streamlines, it therefore follows that a smaller $r_{\rm in}$ for a
given $L/L_{\rm Edd}$ generates more near-IR emission because of the
higher dust temperatures (K12).  However, the opening angle for the
wind becomes quite large, $\theta_{\rm open}=86^\circ$ (see
Fig.~\ref{fig:open}).  If this is the cause of the luminosity
dependence of the 3--5~\micron\ emission, this implies some mechanism
for decreasing the dusty wind launching radius as a function of
luminosity which is challenging to arrange.  We therefore prefer the
geometric explanation above.

\subsection{Images of the Wind}

As luminous quasars are typically farther away and therefore fainter
and of smaller angular size than nearby Seyferts, they are beyond the
reach of the current generation of interferometers.  Therefore we can
only compare our simulated model structure to observations of the
nearest Seyfert galaxies, which are by nature considerably less
luminous (because of lower $M_{\rm BH}$ masses or lower Eddington
ratios or both). A simulated high-resolution image of the wind at
9.5~$\mu$m heated by the central continuum is shown in
Fig.~\ref{fig:image}.  The hourglass shape is characteristic of MHD
winds, and reminiscent of the outflows seen in young stars. It
illustrates clearly that the base of the wind and the inner
streamlines dominate the emission at this wavelength; the brightest
9.5~$\mu$m emission regions would correspond to the largest extinction
values in the optical.  Intriguingly, 8 to 13~$\mu$m interferometric
observations have led \citet{hoenig+12} to infer a similar structure
and characteristic size scale ($\sim2$~pc) for the dust distribution
in the Seyfert~2 galaxy NGC~424.  The largest compilation of
mid-infrared interferometry to date shows that a large fraction of
this emission is unresolved, or at scales of $<0.1$ to $<10$~pc for a
sample of 23 local Seyfert nuclei { \citep{burtscher+13}}.  
Reverberation-mapping using the
response of the near-IR continuum to changes in the optical continuum
of a several Seyfert galaxies has placed the inner `torus' radius
beyond the broad-line region and consistent with the expected
dust-sublimation radius \citep{kishimoto+13}.  However, more direct
comparisons of the dusty wind models with observations await the
light-collecting power and angular resolution
($\sim0.001^{\prime\prime}$ at $2\micron$) expected in the era of 30-m
telescopes.

%------------------------------------------------------------
\begin{figure}
\center
\includegraphics[width=\linewidth]{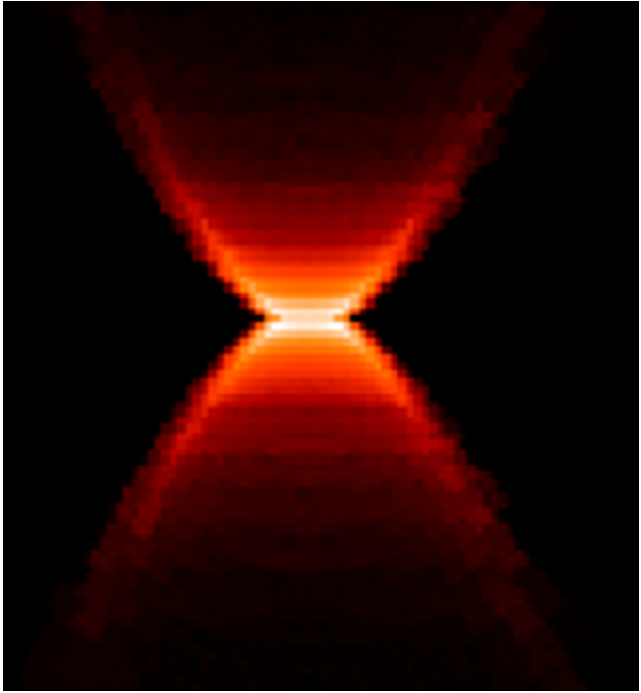}
\caption{\label{fig:image} 
A simulated high-resolution 9.5~$\mu$m image of the fiducial model
generated by {\sc mc3d} \citep{wolf03}.  The hot ``disk''-shaped
structure is approximately 2~pc in radius.  The bulk of the emission
comes from the base of the wind and the inner streamlines heated by
the accretion disk continuum thus giving the MHD wind its
characteristic hour-glass shape.}  \end{figure}

%------------------------------------------------------------

\subsection{Constraints on Organized Magnetic Fields}

The success of the dusty wind model that we have presented is
predicated on the existence of relatively strong, organized magnetic
fields that thread the outer accretion disk of quasars.  Certainly,
radio-loud AGNs have such fields in the inner disk; radio synchrotron
emission from jets requires them.  The picture for radio-quiet AGNs is
less clear.  There is some observational evidence for ordered magnetic
fields within a few parsecs of the central engine of local AGNs as
detected in near-infrared polarimetry.  \citet{Lopez+13} measured
polarization of a few percent in the $K_{\rm n}$ band for the Type~2
AGN IC~5063.  They interpreted the polarization as being due to
dichroic absorption of dust grains aligned by magnetic fields in the
nuclear region.  The polarization properties thus constrain the
$B-$field in the plane of the sky to be 10--130~mG.  However, their
angular resolution is significantly larger ($\sim260$~pc) than the
torus region, and thus the field may be associated with the
interstellar medium of the host galaxy near the nucleus, rather than
the torus itself.  Similarly, from an analysis of Zeeman splitting in
OH megamaser emission in ultra-luminous infrared galaxies,
\citet{robishaw+08} estimate magnetic fields strengths of mG are
present throughout the interstellar medium in these starbursting
galaxies.  On smaller scales (a few tenths of a parsec), the lack of
detections of circular or linear polarization in measurements of the
{\sc H$_2$0} molecular masers in Type~2 AGNs gives upper limits of
10--100~mG for the magnetic fields in the masing region
\citep[][]{modjaz+05,vlemmings+07,mccallum+07}, while our fiducial
model has a field strength at $r_{\rm in}$ near the disk surface of
$\sim0.2$G.  { $B$-field strengths in our $\kappa=0.12$ model are a
factor of four smaller at 50~mG near the disk surface, and therefore}
still viable given these empirical constraints.  Furthermore, the low
luminosity Type~2 AGNs being studied (NGC~4258, NGC~3079, and the
Circinus galaxy) may have central regions that are quite different
from Type~1 quasars, { and so additional empirical constraints on
poloidal $B$-field strengths within the inner parsecs of quasars would be
valuable.}

\section{Summary and Conclusions}

We explore the properties of the wind models first presented in K12,
and extract key parameters including the covering fraction of the
optically thick wind, the column densities as a function of opening
angle of several ions, and the distribution of extinction values for
Type~1 quasars.  In addition, we explore the
effect of decreasing the strength of the poloidal magnetic field and
increasing the column density of the dust-free shielding gas interior
to the dusty wind. Altering key input parameters of the model allows
us to explore the effects of $M_{\rm BH}$, $L/L_{\rm Edd}$, the shape
of the ionizing continuum, the mass-to-magnetic flux ratio, and the
column density of the shielding gas on the wind.  We find the
following empirical consequences of this model:

\begin{itemize}

\item{Using our definition of $\tau<5$ at H$\alpha$ to define
broad-line quasars, the dusty wind model predicts a Type~1 fraction of
$\sim65$\% for our fiducial model (Table~\ref{tab1} and
Fig.~\ref{fig:open}).  Type~2 quasars in our model have line-of-sight
column densities of $N_{\rm H}\gtrsim10^{22}$~\cmsq\
(Fig.~\ref{fig:columns}).  Both of these values are roughly consistent
with observed Type~1 fractions and line-of-sight column densities
found in optical and X-ray surveys of luminous AGNs { and from
mid-IR SED analyses}.}

\item{While the presence of a UV line-driven wind as seen in broad
absorption line quasars apparently requires a blue UV continuum and
weak X-ray emission (either as a result of absorption or intrinsic
weakness), the dust-driven wind is much less sensitive to the {\em
shape} of the continuum.  Mid-IR continuum emission might therefore be
quite similar among classes of quasars with a range of ionizing SED
shapes, and consequently a more accurate measure of bolometric
luminosity.}

\item{We present predictions for the cumulative distribution of
$E(B-V)$ values for quasars (Fig.~\ref{fig:cum_dist}).  The
normalization of these curves depends primarily on $\kappa$, the
mass-to-magnetic flux ratio, with higher values of $\kappa$ (weaker
$B$-fields) resulting in slightly larger opening angles, and
significantly less dust extinction over a wide range of opening
angles.  This is a consequence of radiation pressure dominating over
magnetic driving which results in more radial outflows.  The effect of
changing $\kappa$ has a larger effect on the $E(B-V)$ distribution
than changing $L/L_{\rm Edd}$ within the range we considered.}

\item{The column density through the wind drops rapidly with height
from the disk causing the ionization parameter of the wind to increase
dramatically.  As a consequence, high ionization atomic species are
present at a large range of inclination angles.  The modeled values
for the column densities of {O}~{\sc viii} through the wind are
$\sim10^{16-18}$~\cmsq\ depending on the properties of the ionizing
continuum and the value of $\kappa$ (Fig.~\ref{fig:columns}). 
{ This range is low for the strongest 
X-ray warm absorbers in the best-studied
local Seyfert galaxies, but consistent with more typical warm absorbers.}
Improved measurements of the profiles of warm
absorber features with next generation X-ray observatories may
determine if the dusty wind model presented here remains a viable
explanation for X-ray warm absorbers.}

\item{The hot dust emission present in the near-IR SEDs of luminous
quasars could be associated with the innermost surface of the dusty
wind, where grains with higher sublimation temperatures than silicates
can survive.  We predict that the strength of this emission feature
will correlate with $L/L_{\rm Edd}$ as higher $L/L_{\rm Edd}$ flows
are more radial, and the cross-section of the inner streamline visible
to the observer at typical inclination angles becomes larger.  Scatter
could be introduced in this correlation by a distribution of magnetic
field strengths at given values of $L/L_{\rm Edd}$.}

\item{MHD+radiation wind driven models of the AGN `torus' predict
hour-glass shapes for the mid-IR emission which may be routinely
detectable in nearby AGN with improvements in near-to-mid IR
interferometry and the next generation of 30-m-class telescopes.  The
shape of the innermost streamline (which generates the bulk of the
emission) is most sensitive to the mass-to-magnetic flux ratio, and
secondly to $L/L_{\rm Edd}$.}

\end{itemize}

Getting the dusty-wind model to the point where it can plausibly
account for the detailed structure in near-to-mid-infrared quasar SEDs
requires additional refinement.  In particular, the current
single-zone dust structure is not realistic, and the dust composition
and grain-size distribution likely varies from that of the Milky Way
ISM as assumed in our models.  The need for different dust properties
for quasar environs is already implied by observations indicating
that the shape of the extinction curve to most quasars is more
consistent with SMC-type dust \citep[e.g.,][]{coleman+14}.

While clouds are not likely to persist in the dynamic environment
around a supermassive black hole \citep[e.g.,][]{schartmann}, it is
also unlikely that any wind structure is completely smooth. Including
inhomogeneities in the wind is therefore an obvious next step, in
particular to more realistically predict ionic column densities.

More generally, the viability of the dusty MHD disk wind paradigm
requires an investigation into whether relatively strong ($\gtrsim50$ mG),
large-scale poloidal magnetic fields persist within the central
parsecs of quasar host galaxies.  Certainly quasars with jets have
large-scale magnetic fields; radio synchrotron emission is observed.
In the near future, the available sensitivity and spatial resolution
of {\it ALMA} and the Jansky VLA are promising for placing meaningful
constraints on the strengths of organized fields within the central
few parsecs of radio-quiet active galactic nuclei.

\section*{Acknowledgments} This work was supported by the Natural
Science and Engineering Research Council of Canada and the Ontario
Early Researcher Award Program.  We thank R.~P. Deo for his
contributions, and J. Cami, M. Houde, K. Leighly, and G. Richards for helpful
discussions.  This work was supported in part by the National Science
Foundation under Grant No. PHYS-1066293 and the hospitality of the
Aspen Center for Physics.  { We thank the anonymous referee for
thoughtful comments that improved the presentation of the manuscript.}

%-------------------------------------
\bibliographystyle{mn2e}
\bibliography{dustywinds2}
%-------------------------------------

\label{lastpage}

\end{document}

%% file: tab1.tex
\begin{table*}
 \centering
% \begin{minipage}{140mm}
  \caption{Properties of Dusty Wind Models. All models have \aox=--1.6 and $M_{\rm BH}=10^{8}$~\msun\ except for model 2 (fid\_mbh8p7) with 
$M_{\rm BH}=5\times10^8$~\msun.
\label{tab1}}
  \begin{tabular}{@{}lccc|ccccc@{}}
  \hline
   Model Name& $\kappa^a$ & $L/L_{\rm Edd}$ & $N_{\rm H,shield}^b$ & $r_{\rm in}^c$ &  \multicolumn{2}{c}{$\theta_{\rm open}^d$ (degrees)} & $\dot{m}_{\rm outflow}^e$  & $v_{\rm term}$  \\
         &          &                 & $\log({\rm cm^{-2}})$ &  ($10^{18}$~cm) &  $\tau_{\rm H\alpha}=3$ & $\tau_{\rm H\alpha}=5$          &  ($M_{\odot}$~yr$^{-1}$)    &($10^3$~\kms)              \\
 \hline
1. fiducial (A)$^g$ 	   & 0.03 & 0.10 & 18    & 4.57 & 62.3 & 69.4  & 1.20    & 2.94 \\
2. fid\_mbh8p7 (B)$^g$      & 0.03 & 0.10 & 18   & 10.6 & 62.2 & 69.3  & 3.80	 & 5.00 \\
3. fid\_edd001 (C)$^g$      & 0.03 & 0.01 & 18   & 1.60 & 60.4 & 67.3  & 0.99    & 1.90 \\
4. sg18\_rsublg\_k04         & 0.04 & 0.10 & 18  & 4.57 & 63.7 & 70.0  & 1.21    & 2.89 \\
5. sg18\_rsublg\_k06         & 0.06 & 0.10 & 18  & 4.57 & 66.9 & 71.7  & 1.21    & 2.89 \\
6. sg18\_rsublg\_k12         & 0.12 & 0.10 & 18  & 4.57 & 74.0 & 76.5  & 1.22    & 2.77 \\
7. sg18\_rsubsm\_k04\_edd05  & 0.04 & 0.05 & 18  & 3.23 & 63.0 & 69.1  & 1.07    & 2.81 \\
8. sg18\_rsubsm\_k06\_edd05  & 0.06 & 0.05 & 18  & 3.23 & 66.5 & 71.2  & 1.07    & 2.78 \\
9. sg18\_rsubsm\_k12\_edd05  & 0.12 & 0.05 & 18  & 3.23 & 73.8 & 76.4  & 1.08    & 2.77 \\
10. sg19\_rsub4pt570        & 0.03 & 0.10 & 19   & 4.57 & 62.3 & 69.4  & 1.20    & 2.94 \\
11. sg23\_rsub4pt480        & 0.03 & 0.10 & 23   & 4.48 & 62.3 & 69.4  & 1.19    & 2.95 \\
12. sg24\_rsub4pt320        & 0.03 & 0.10 & 24   & 4.32 & 62.3 & 69.4  & 1.18    & 2.97 \\
13. sg25\_rsub3pt900        & 0.03 & 0.10 & 25   & 3.90 & 62.3 & 69.4  & 1.13    & 3.03 \\
\hline
\multicolumn{9}{p{0.9\textwidth}}{$^a$Input mass-to-magnetic flux ratio.  
A smaller value indicates a stronger poloidal $B$-field.
$^b$Input column density of the shielding (dust-free) gas at the base of the wind.
$^c$The dust sublimation radius at the base of the wind calculated based on the input parameters.  This defines the launching radius of the innermost wind streamline.
$^d$The opening angle of the dusty wind as defined for $\tau=3$ and 5 at H$\alpha\lambda6563$.
$^e$The mass-outflow rate of the dusty wind.
$^f$The terminal velocity of the dusty wind.
$^g$Letters in parentheses refer to the model labels in Figure~\ref{fig:streamlines}.  
}
\end{tabular}
%\caption{$^a$Input mass-to-magnetic flux ratio.  A smaller value
%indicates a stronger poloidal $B$-field.  $^b$Input column density of
%the shielding (dust-free) gas at the base of the wind.  $^c$The dust
%sublimation radius at the base of the wind calculated based on the
%input parameters.  $^d$The opening angle of the dusty wind as defined
%for $\tau=3$ and 5 at $H\alpha\lambda6564$.  $^e$The terminal velocity
%of the dusty wind.  $^f$The mass-outflow rate of the dusty wind.
%$^g$Letters in parentheses refer to the model labels in
%Figure~\ref{fig:streamlines}.  All models have $M_{\rm
%BH}=10^{8}$~\msun\ except for model fid\_mbh8p7 with $M_{\rm
%BH}=5\times10^8$~\msun.}  
%\end{tabular} 
%\end{minipage} 
\end{table*}